\documentclass[aps,prl,twocolumn,superscriptaddress]{revtex4}
\usepackage{amsmath}
\usepackage{graphicx}
\usepackage[section]{placeins}

\newcommand{\eq}{\begin{equation}}
\newcommand{\fine}{\end{equation}}
\newcommand{\ket}[1]{|#1\rangle}

\newcommand{\eac}{\'e}

\begin{document}

\title{Quantum information transfer from spin to orbital angular momentum of photons}
\author{Eleonora Nagali}

\affiliation{Dipartimento di Fisica dell'Universit\`{a} ``La
Sapienza'' and Consorzio Nazionale Interuniversitario per le Scienze
Fisiche della Materia, Roma 00185, Italy}

\author{Fabio Sciarrino}

\affiliation{Dipartimento di Fisica dell'Universit\`{a} ``La
Sapienza'' and Consorzio Nazionale Interuniversitario per le Scienze
Fisiche della Materia, Roma 00185, Italy}

\author{Francesco De Martini}

\email{francesco.demartini@uniroma1.it}

\affiliation{Dipartimento di Fisica dell'Universit\`{a} ``La
Sapienza'' and Consorzio Nazionale Interuniversitario per le Scienze
Fisiche della Materia, Roma 00185, Italy}

\affiliation{Accademia Nazionale dei Lincei, via della Lungara 10,
Roma 00165, Italy}

\author{Lorenzo Marrucci}

\email{lorenzo.marrucci@na.infn.it}

\affiliation{Dipartimento di Scienze Fisiche, Universit\`{a} di
Napoli ``Federico II'', Compl.\ Univ.\ di Monte S. Angelo, 80126
Napoli, Italy}

\affiliation{ CNR-INFM Coherentia, Compl.\ Univ.\ di Monte S.
Angelo, 80126 Napoli, Italy}

\author{Bruno Piccirillo}

\affiliation{Dipartimento di Scienze Fisiche, Universit\`{a} di
Napoli ``Federico II'', Compl.\ Univ.\ di Monte S. Angelo, 80126
Napoli, Italy}

\affiliation{Consorzio Nazionale Interuniversitario per le Scienze
Fisiche della Materia, Napoli}

\author{Ebrahim Karimi}

\affiliation{Dipartimento di Scienze Fisiche, Universit\`{a} di
Napoli ``Federico II'', Compl.\ Univ.\ di Monte S. Angelo, 80126
Napoli, Italy}

\author{Enrico Santamato}

\affiliation{Dipartimento di Scienze Fisiche, Universit\`{a} di
Napoli ``Federico II'', Compl.\ Univ.\ di Monte S. Angelo, 80126
Napoli, Italy}

\affiliation{Consorzio Nazionale Interuniversitario per le Scienze
Fisiche della Materia, Napoli}

\begin{abstract}
The optical ``spin-orbit'' coupling occurring in a suitably
patterned nonuniform birefringent plate known as `q-plate' allows
entangling the polarization of a single photon with its orbital
angular momentum (OAM). This process, in turn, can be exploited for
building a bidirectional ``spin-OAM interface'', capable of transposing
the quantum information from the spin to the OAM degree of freedom
of photons and \textit{vice versa}. Here, we experimentally demonstrate this
process by single-photon quantum tomographic analysis. Moreover,
we show that two-photon quantum correlations such as those resulting
from coalescence interference can be successfully transferred into the OAM degree of freedom.
\end{abstract}

\maketitle

 In the last few decades, quantum optics has allowed the
implementation of a variety of quantum-information protocols.
However, the standard information encoding based on the
two-dimensional quantum space of photon polarizations (or ``spin''
angular momentum) imposes significant limitations to the protocols
that may be implemented. To overcome such limitations, more
recently the orbital angular momentum (OAM) of light, related to
the photon's transverse-mode spatial structure \cite{Alle92}, has
been recognized as a new promising resource, allowing the
implementation of a higher-dimensional quantum space, or a
``qu-dit'',  encoded in a single photon
\cite{Moli07,Fran08,Aiel05}. So far, the generation of
OAM-entangled photon pairs has been carried out by exploiting the
process of parametric down-conversion \cite{Mair01,Moli03} and has
also been utilized in few quantum information protocols
\cite{Leach02,Vazi03,Barr05,Barr08}. Despite these successes, the
optical tools for generating and controlling the OAM photon states
(computer-generated holograms, Dove's prisms, cylindrical lens
, etc.) are rather limited. A convenient way to
coherently ``interface'' the OAM degree of freedom of photons with
the more easily manipulated spin/polarization one has been missing
so far. In this context, the recent invention of an optical
device, the so-called ``q-plate'' (QP), that couples the photon
spin to its orbital angular momentum opens up a number of new
possibilities \cite{Marr06}. In this work we show, both
theoretically and experimentally, that this optical ``spin-orbit''
coupling can be exploited as an effective coherent bidirectional interface
 between polarization and orbital angular
momentum in the quantum regime. The QP also enables the efficient
generation of single-photon states in which the OAM and polarization
degrees of freedom are entangled. Furthermore, we show that photon-photon quantum correlations of a biphoton \cite{Bogd04}
can be transferred from the spin to the OAM degree of freedom.\\
\begin{figure}[t]
\centering
\includegraphics[scale=.33]{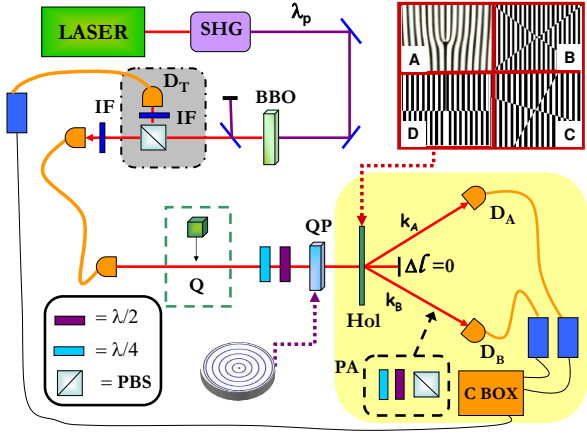}
\caption{A Ti:Sa
mode-locked laser converted by second harmonic generation (SHG)
into a beam with wavelength $\lambda_p=397.5$nm. This field pumps
a nonlinear crystal of $\beta$-barium borate (BBO) which emits a
single-mode biphoton state with $H$ and $V$ polarizations and
$\lambda=795$nm, filtered by the interference filter (IF) with
$\Delta\lambda=6$nm and then coupled to a single mode fiber
\cite{DeMa05}. The gray dot-dashed box has been optionally
inserted to prepare a single photon state triggered by detector
$D_{T}$. Birefringent quartz crystals (Q) having different
thicknesses were used to introduce a controlled temporal delay
between the two photons. After setting the input polarization by
means of a suitably oriented quarter-wave plate, the photons were
sent through the q-plate (QP) and the output OAM states were
analyzed with the help of a hologram (Hol). In OAM-to-spin
conversion experiments, Hol and QP were interchanged. To measure
(or prepare) OAM states in the basis $l=\pm 2$, a double-fork
hologram has been used (inset $A$), so that the OAM state of the
first diffracted modes is shifted by $\Delta l=\pm2$, while the
undiffracted 0-order beam has $\Delta l=0$. The photons on the
first diffracted modes are then coupled to single mode fibers
which select output states with $l=0$ and convey them to the
detectors $D_A$ and $D_B$. Hence, the detection of a photon in
$D_A$ ($D_B$) corresponds to a photon incident on the hologram
with OAM $l=+2$ ($l=-2$). The 1st-order diffraction efficiency of
the hologram was $\sim 10\%$. The measurement (or preparation) of
OAM in superposition states has been realized by adopting the
other holograms shown in the inset. (The hologram $B$ refers to $\ket{d_+}_l$, $C$ to
$\ket{d_R}_l$, $D$ to $\ket{d_-}_l$, $\ket{d_L}_l$ was also analyzed by hologram $C$ after reversing its orientation"). Due to reflection losses,
the transmittance of the q-plate is $T\sim 0.90$.}
\end{figure}
The QP is a birefringent slab having a suitably patterned
transverse optical axis, with a topological singularity at its
center. Here, we consider a QP with ``charge'' of the singularity
$q=1$ and uniform birefringent retardation $\delta=\pi$. Such QP
modifies the OAM quantum number $l$ (in units of $\hbar$) of a
light beam crossing it, imposing a variation $\Delta{l}={\pm}2$
whose sign depends on the input polarization, positive for
left-circular and negative for right-circular. The helicity of the
output circular polarization is also inverted, i.e. the optical
spin is flipped \cite{Calv07}. Let us now rephrase this behavior
in a quantum formalism suitable for describing multi-photon
states. Let $\widehat{a}_{j,l}^{\dagger}$ be the operator creating
a photon in the polarization state $\overrightarrow{\pi}_{j}$
(where $\overrightarrow{\pi}_{H}$ and $\overrightarrow{\pi}_{V}$
stand for horizontal and vertical linear polarizations,
respectively) and with the OAM value $l$. For simplicity, we omit
here the radial quantum number which would be needed for a
complete characterization of transverse orbital states, as it will
play no significant role in the following \cite{footnote_radial}.
The overall dynamics induced by the QP can be then described by
the transformations
$\widehat{a}_{L,0}^{\dagger}\Rightarrow\widehat{a}_{R,2}^{\dagger}$
and $\widehat{a}_{R,0}^{\dagger} \Rightarrow
\widehat{a}_{L,-2}^{\dagger}$, with
$\overrightarrow{\pi}_{L}=2^{-1/2}\left(
\overrightarrow{\pi}_{H}+i \overrightarrow{\pi}_{V}\right) $ and
$\overrightarrow{\pi}_{R}= \overrightarrow{\pi}_{L}^{\perp}$
standing for left and right
circular polarizations, respectively.\\
The previous relations describe the coupling of the OAM $l$ and the polarization $\pi$
degrees of freedom taking place in the QP.
Interestingly, this property can be exploited to generate single-particle entanglement of $\pi$ and
$l$ degrees of freedom:
\begin{equation}
\left.
\begin{array}{c}
\left| H\right\rangle _{\pi }\left| 0\right\rangle _{l} \\
\left| V\right\rangle _{\pi }\left| 0\right\rangle _{l}
\end{array}
\right\} \overset{QP}{\rightarrow
}\frac{1}{\sqrt{2}}(\ket{L}_{\pi}\ket{-2}_l\pm\ket{R}_{\pi}\ket{+2}_l)
\label{entangled}
\end{equation}
This is an entangled state between two qubits encoded in different
degrees of freedom. In particular $\{\ket{+2}_l,\ket{-2}_l\}$ is
the basis for the OAM qubit which lies in the $|l|=2$ subspace of
the infinite dimensional Hilbert space of orbital angular
momentum. As a first experimental step, we set out to verify how
accurately the real QP device performs these transformations in
the single photon regime (see Fig.1 for experimental details).
First, the QP conversion efficiency $\eta$ from the input
TEM$_{00}$ to the $l=\pm 2$ modes has been estimated through the
coupling efficiency with the single mode fiber. We find $\eta
\simeq 85\%$, ascribed to light scattering, radial mode residual mismatch, and imperfect tuning of the QP birefringent retardation $\delta$ \cite{Marr06,eff} (the
unconverted component remains $l=0$ and is therefore filtered
out).
\begin{figure}[t!!]
\centering
\includegraphics[scale=.25]{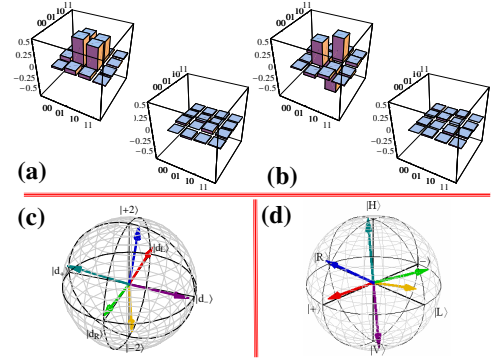}
\caption{\textbf{(a-b)}Experimental density matrices for the
single photon entangled state. The computational values $\{0,1\}$
are associated to the $\{\ket{R},\ket{L}\}$ polarization states,
and to $\{\ket{+2},\ket{-2}\}$ for the orbital angular momentum
$l$ for the first and the second qubit, respectively. The incoming
state on the QP is (\textbf{a}) $\ket{H}_{\pi}\ket{0}_l$,and
(\textbf{b}) $\ket{V}_{\pi}\ket{0}_l$. The average experimental
concurrence is $C=(0.95\pm0.02)$. \textbf{(c-d)}Experimental Poincar{\eac}
sphere both for the OAM \textbf{(c)} and $\pi$ \textbf{(d)} degrees
of freedom obtained after the $\pi\rightarrow l$ and the $l\rightarrow\pi$ transferrer, respectively. Experimentally we carried out single-qubit tomography
 to determine the Stoke's parameters for the $\pi$ and
their analogous for $l$ degrees of freedom. The mean fidelity
values are \textbf{(c)} $F=(98\pm1)\%$ and  \textbf{(d)}
$F=(97\pm1)\%$.}
\end{figure}
Next, in order to assess the coherence of the transformations in
Eq. (1), single photons in the states $\ket{H}_{\pi}\ket{0}_l$ or
$\ket{V}_{\pi}\ket{0}_l$ were used as input in the QP. We analyzed
the output state through a double-fork hologram and a
circular-polarization analysis setup along the two diffracted
modes: the intensity of the $\pi_R$ ($\pi_L$) polarization
component in the mode corresponding to $l=+2$ ($l=-2$) was
measured to be equal to 99.8\% (99.6\%) of the total, with a high
agreement with theory. To demonstrate the realization of the pure
states given in Eq.(\ref{entangled}), a complete single-photon
two-qubit quantum state tomography has been carried out,
performing measurements both in $\pi$ and $l$ degrees of freedom.
Besides the normal $\{\ket{+2}_l,\ket{-2}_l\}$ OAM basis,
measurements were carried out in the two superposition bases
$\{\ket{d_+}_l,\ket{d_-}_l\}$ and $\{\ket{d_R}_l,\ket{d_L}_l\}$,
where
$\ket{d_{\pm}}_l=\frac{1}{\sqrt{2}}(\ket{+2}_l\pm\ket{-2}_l)$ and
$\ket{d_{L,R}}_l=\frac{1}{\sqrt{2}}(\ket{+2}_l\pm i\ket{-2}_l)$.
The OAM degree of freedom was analyzed in these bases by means of
different computer-generated holograms, reported in the
inset in Fig.(1) \cite{Lang04}. The experimental results are in
high agreement with theory, as shown in
Fig.(2-\textbf{a,b}).

 Due to its pecularities, the q-plate
provides a convenient way to "interface" the photon OAM
with the more easily manipulated spin degree of freedom. Hence as
next step we show that such interface can be considered as a
quantum "transferrer" device, which allows to transfer coherently
the quantum information from the polarization $\pi$ to the OAM $l$ degree
of freedom, and \textit{vice versa}. Such processes can be
formally expressed as : $ \ket{\varphi}_{\pi}\ket{0}_l\leftrightarrow \ket{H}_{\pi}\ket{\varphi}_l$
where $\ket{\varphi}_{\pi}=(\alpha\ket{H}+\beta\ket{V})_{\pi}$ and
$\ket{\varphi}_l=(\alpha\ket{+2}+\beta\ket{-2})_l$. Here we
demonstrate a probabilistic conversion (with probability
$p\cong50\%$), since some output state contributions are
discarded. By extending the present scheme, it is possible to
achieve a complete
deterministic information conversion \cite{ext}.\\
\textbf{I - Transferrer $\pi\rightarrow l$}. Through a waveplate
$\lambda/4$ and a QP, $\pi$ and $l$ become entangled (Fig.
3-\textbf{I-a}). Then, the information contained in the polarization
is erased by inserting a polarizing beam-splitter (PBS). This
process has been experimentally verified for a set of maximally
polarized states. Any $\pi$ input state, represented by a vector in
the Poincar{\eac} sphere, has been converted in another vector in
the OAM ``Poincar{\eac} sphere'' (see Fig.2-\textbf{c}), determined by carrying out a quantum state tomography of the OAM
state. The results
demonstrate a high fidelity of the $\pi\rightarrow l$
transformation. An application of this transferrer is shown
in Fig.(3-\textbf{I-b}). The initial information encoded in an input
state $\ket{\varphi}_{\pi}\ket{0}_l$ is coherentely transferred to the
OAM. New information can then be stored in the polarization degree
of freedom, thus allowing the encoding of two qubits of information
in the same photon.
\begin{figure}[b]
\centering
\includegraphics[scale=.31]{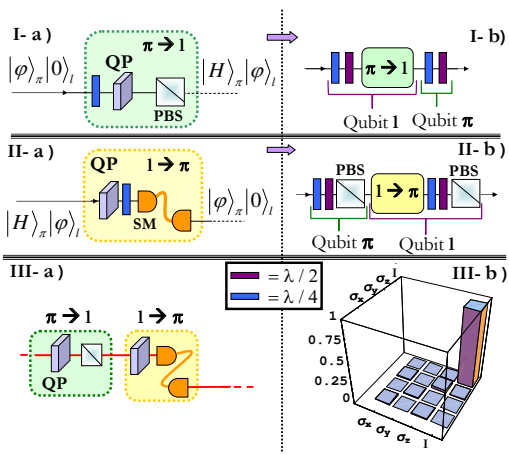}
\caption{(\textbf{I-II})Schematic representation of the two transferrer devices based
on the q-plate (QP).\textbf{III} a) Double transfer $\pi\rightarrow l\rightarrow \pi$, b) Experimental real part of the matrix $\chi$ represented in the Pauli operator basis, the imaginary elements are $\approx 0$ . }
\end{figure}
\textbf{II - Transferrer $l\rightarrow \pi$.} Let us show that the
coupling between the spin and the orbital degree of freedom
in the q-plate is bidirectional (Fig.3-\textbf{II-a}). The
input state $\ket{H}_{\pi}\ket{\varphi}_{l}$ is sent through a QP
and a $\lambda/4$ waveplate (Fig.\ 3-\textbf{II-a}). By inserting
a single mode fiber on the output, only the states with $l=0$ are
efficiently coupled, leading to a probabilistic process. After the
fiber, the state $\ket{\varphi}_{\pi}\ket{0}_{l}$ has been
obtained, which demonstrates the successful transfer from the OAM
degree of freedom to the polarization one. Analogously to the
previous analysis, the experimental Poincar{\eac} sphere vectors obtained
after the $l\rightarrow\pi$ conversion are reported in
Fig.(2-\textbf{d}). Such transferrer can be for example exploited
within a measurement apparatus for analyzing both $\pi$ and $l$
degrees of freedom without using holograms, with an advantage in
terms of efficiency (Fig.(3-\textbf{II-b})) and flexibility, since
there is no need to change the hologram for each state to be
analyzed. Finally we proved the forward-backward double
transfer $\pi\rightarrow l\rightarrow\pi$, by implementing both
the previous transferrers together, as shown in
Fig.(3-\textbf{III-a}). 
We carried out the
quantum process tomography  \cite{Niel00} proving that the qubit state quantum information is coherently preserved in the whole process involving two separate QPs. Fig.(3-\textbf{III-b}) reports the reconstructed $\chi$ matrix representation of the overall process which exhibits a fidelity, i.e. overlap, with the identity map equal to $F=0.950\pm0.015$. This bidirectional spin-OAM interface allows the extension to OAM of many protocols currently
only possible with polarization. For example, the realization of a
two-photon \emph{C-NOT} gate for OAM states can be obtained by
exploiting a \emph{C-NOT} for $\pi$ states. As a further
consideration, the QP allows implementing a fast OAM switcher
device, by modulating the polarization degree of freedom through a
Pockel's cell and then transferring such modulation to the OAM.

 It
is natural to ask if the QP is able to preserve the photon-photon
correlations too. This would also be a crucial test of the QP
potential in the quantum optics field, as multi-photon
correlations are extremely sensitive to optical quality issues.
Consider first the case of two independent linearly
polarized photons, one horizontal and the other vertical, going
through the QP: each will undergo the QP transformation given in
Eq.(\ref{entangled}) and the two outgoing photons will end up
having opposite OAM values 50\% of the times. If, however, the two
photons are undistinguishable, except for their polarization, the
field state at the QP input side is $\ket{\Psi_{in}} =
\ket{1}_{H,0}\ket{1}_{V,0}$, where we are now using a photon-number
ket notation for our multi-photon quantum states. This input state
can be rewritten in the circular-polarization basis as
$\ket{\Psi_{in}} = \frac{1}{\sqrt{2
}}(\ket{2}_{R,0}\ket{0}_{L,0}-\ket{0}_{R,0}\ket{2} _{L,0})$. After
the QP, this state evolves into
\begin{equation}
\ket{\Psi _{out}} =\frac{1}{\sqrt{2}}(\ket{2}_{L,-2}\ket{0}_{R,2}-\ket{0}_{L,-2}\ket{2}_{R,2})
\label{oumandel},
\end{equation}
in which only photons carrying parallel OAM, either $+2$ or $-2$ are
found. Hence, the QP action can be again interpreted as a
mode converter, coherently transferring the two-photon
quantum correlation from the spin degree of freedom to the OAM one.
\begin{figure}[h]
\centering
\includegraphics[scale=.3]{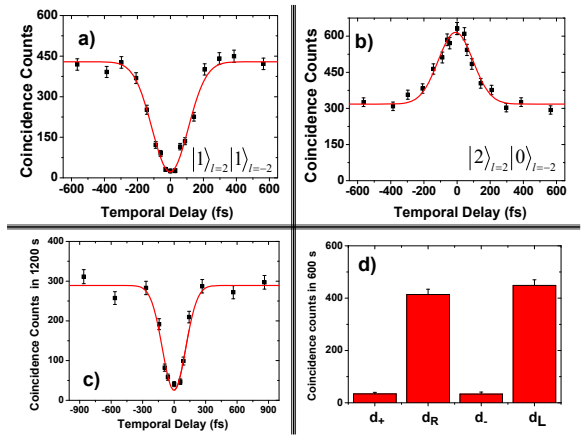}
\caption{(\textbf{a}) Coincidence counts between $[D_{A},D_{B}]$
versus the temporal delay $t_d$, for the state $\ket{\Psi _{out}}$.
The continuous line shows the best fit of the experimental data.
(\textbf{b}) Coincidence counts $[D_{A},D_{A^{\prime}}]$ versus
$t_d$. \textbf{(c)} Coincidence counts $[D_{A},D_{B}]$ versus $t_d$
($V=(0.91\pm 0.01)$) for the state $\ket{\Phi _{out}}$. \textbf{(d)}
Coincidence counts $[D_{A},D_{A^{\prime}}]$ in different OAM bases.}
\end{figure}
The QP is then able to generate a multi-photon state having
non-classical correlations in OAM within a single longitudinal
mode. The experimental setup is shown in Fig.1 (with the gray box removed). The coincidence
between $[D_{A},D_{B}]$ detects the state contribution of
$\ket{1}_{l=2}\ket{1}_{l=-2}$. Due to the coalescence
interference, for otherwise identical photons this component is
vanishing in the state outcoming from the QP. In order to study
the transition from the case of classical behaviour (independent
photons) to the case of full quantum interference, a variable
temporal delay $t_d$ between the $H$ and $V$ polarizations in the
state $\ket{\Psi _{in}}$ has been introduced (Q in Fig.1). The
experimental visibility of the quantum interference shown in
Fig.(4-\textbf{a}) is $V_{exp} =(0.95\pm0.02)$. As a further
confirmation, we have measured the contribution of two photons
with $l=+2$ by recording the coincidence counts between two
detectors $[D_{A},D_{A}^{\prime }]$ placed on the output modes of
a fiber-based beam splitter inserted on the same $k_{A}$
diffracted mode (not shown in figure). Theoretically, the
coalescence of the two photons should lead to a coincidence
enhancement by a factor $\Gamma=2$, and experimentally we found
$\Gamma _{exp}=(1.94\pm 0.02)$. For the sake of completeness, we verified that even after erasing
all information still contained in the polarization degree of
freedom, the final state is still coherent and exhibits the same
non-classical photon correlations in OAM. To this purpose, we let
both the two photons pass through a horizontal linear polarizer
set in a common state $H$. We verified again the coalescence of
the photons by a measurement similar to the previous one: Fig.(4-\textbf{c}). To verify
that we really obtain a coherent state  and not a statistical
mixture having similar OAM correlation properties, we measured the coherence between the two contributions with opposite
OAM states. This was accomplished by analyzing the photons in the
other OAM bases already discussed above. Therefore, our coherence
verification is actually turned into a measurement of two photons
in the same OAM state $(\ket{d_+},\ket{d_-},\ket{d_R},\ket{d_L})$.
As expected, for $(\ket{d_+},\ket{d_-})$ the events of two photons
with the same orbital states are strongly suppressed, while for
$(\ket{d_R},\ket{d_L})$ they are doubled, with an overall
correlation visibility $V=(0.86\pm0.02)$ (Fig.\ 4(\textbf{d})).
This shows that QPs transfer not only single-photon
information between polarization and OAM, but also
multiphoton-encoded information (e.g. biphotons).

 In conclusion, we have shown that the q-plate device
can be used as a coherent and bidirectional quantum interface
allowing the transfer of quantum information between the
polarization and the orbital angular momentum degrees of freedom
of the photon, both in the case of single photon states and of
two-photon correlated states. The results
reported here show that this can be a useful tool for exploiting the
OAM degree of freedom of photons, in combination with
polarization, as a new resource to implement high-dimensional
quantum information protocols. We thank G. Maiello for assistance
in preparing the q-plates, P. Mataloni and G. Vallone for
useful discussions, and M. Barbieri for useful help on quantum process tomography.


\begin{thebibliography}{99}
\bibitem{Alle92}  L. Allen, et al., \textit{Phys.\ Rev.\ A} \textbf{45}, 8185 (1992)

\bibitem{Moli07}  G. Molina-Terriza, et al., \textit{Nature Phys.} \textbf{3}, 305 (2007)

\bibitem{Fran08} S. Franke-Arnold, et al., \textit{Laser \& Photon. Rev.}, \textbf{4}, 299313 (2008)

\bibitem{Aiel05} S. S. R. Oemrawsingh, et al., \textit{Phys. Rev. Lett.}, \textbf{95}, 240501 (2005)

\bibitem{Mair01}  A. Mair, et al., \textit{Nature (London)} \textbf{412}, 313 (2001)

\bibitem{Moli03} G. Molina-Terriza, et al., \textit{Opt.\ Commun.} \textbf{228}, 115 (2003)

\bibitem{Leach02} J. Leach et al., \textit{Phys.\ Rev.\ Lett.} \textbf{88}, 257901
(2002).

\bibitem{Vazi03}  A. Vaziri, et al., \textit{Phys.\ Rev.\ Lett.} \textbf{91}, 227902
(2003).

\bibitem{Barr05} J.T. Barreiro, et al., \textit{Phys.\ Rev.\ Lett.} \textbf{95}, 260501
(2005).

\bibitem{Barr08} J.T. Barreiro, et al., \textit{Nature Phys.} \textbf{4}, 282
(2008).

\bibitem{Marr06}  L. Marrucci, et al., \textit{Phys.\ Rev.\ Lett.} \textbf{96},
163905 (2006).

\bibitem{Bogd04} Y. I. Bogdanov, et al., \textit{Phys. Rev. Lett.} \textbf{93}, 230503 (2004).

\bibitem{eff}Since $\delta$ is function of the q-plate thickness,
the wavelenght and the liquid crystal birefringence, it is
possible to optimize the QP efficiency by changing the pressure
acting on the q-plate and the working temperature.

\bibitem{Calv07} G. F. Calvo, et al., \textit{Opt. Lett.} \textbf{32}, 7 (2007)

\bibitem{footnote_radial}
The radial profile generated by the QP, as well as by other
OAM-generating methods, is a
well defined hypergeometric-gaussian mode [see, E. Karimi et al.,
\textit{Opt.\ Lett.} \textbf{32}, 3053 (2007); Karimi et al.,
arXiv0809.4220v1]. In OAM-based quantum optics, when only states
with opposite value of the OAM are involved, the radial profile is
not critical, because it turns out to be independent of the OAM sign and hence it factorizes and does not lead to OAM decoherence. Its role is expected to be more critical when OAM states with different value of $|l|$ are manipulated simultaneously.

\bibitem{DeMa05} F. De Martini, et al., \textit{Prog.\ in Quant.\ Elect.} \textbf{29}, 165 (2005).

\bibitem{Lang04} N.K. Langford, et al., \textit{Phys.\ Rev.\ Lett.} \textbf{93}, 053601 (2004).

\bibitem{ext} By inserting a q-plate on the input mode of a Mach-Zehnder interferometer with a $\sigma_z$ (i.e. a Dove's prism) acting on the OAM in one of its arms, it can be demonstrated that such setup allows to transfer the information from $\pi$ to $l$ deterministically.

\bibitem{Niel00} M. Nielsen and I. Chuang, Quantum Computation and Information, Cambridge (2000).

\end{thebibliography}
\end{document}